\documentclass{iopart}
\usepackage{graphics,graphicx}
\usepackage{subfigure}
\usepackage{iopams}

\def \bsigma {\mbox {\boldmath $\sigma$}}
\def \bxi {\mbox {\boldmath $\xi$}}
\def \bm {\mbox {\boldmath $m$}}
\def \bA {\mbox {\boldmath $A$}}
\def \bI {\mbox {\boldmath $I$}}
\def \bS {\mbox {\boldmath $S$}}

\def \bpsi {\mbox {\boldmath $\psi$}}
\def \bphi {\mbox {\boldmath $\phi$}}
\def \bG {\mbox {\boldmath $G$}}
\def \bC {\mbox {\boldmath $C$}}

\begin{document}

\title[Recurrent neural network model with a synchronous dynamics]
{Symmetric sequence processing in a recurrent neural network model
with a synchronous dynamics}

\author{F L Metz \footnote{Present address:
Instituut voor Theoretische Fysica, Katholieke Universiteit Leuven, Celestijnenlaan
|200D, B-3001 Leuven, Belgium.} and W K Theumann}
\address{Instituto de F\'\i sica, Universidade Federal do Rio
Grande do Sul, Caixa Postal 15051, 91501-970 Porto Alegre, Brazil}
\eads{\mailto{theumann@if.ufrgs.br}, \mailto{fmetz@if.ufrgs.br}}

\date{\today}
\thispagestyle{empty}

\begin{abstract}

The synchronous dynamics and the stationary states of a recurrent
attractor neural network model with competing synapses between
symmetric sequence processing and Hebbian pattern reconstruction is
studied in this work allowing for the presence of a self-interaction
for each unit. Phase diagrams of stationary states are obtained
exhibiting phases of retrieval, symmetric and period-two cyclic
states as well as correlated and frozen-in states, in the absence of
noise. The frozen-in states are destabilised by synaptic noise and
well separated regions of correlated and cyclic states are obtained.
Excitatory or inhibitory self-interactions yield enlarged phases of
fixed-point or cyclic behaviour.

\end{abstract}

\pacs{75.10.Hk, 87.18.Sn, 02.50.-r }
\submitto{\JPA}

\maketitle

\section{Introduction}

The asymptotic stationary states of large recurrent attractor neural
network models trained with sequences of patterns have been studied
some time ago \cite{SK86}-\cite{YYK01} and there has been a recent
revival of interest near the storage saturation of patterns
\cite{AC01}-\cite{CZYZ08}. Besides network models for asymmetric
sequence processing, models with synapses generated by symmetric
sequences competing with pattern reconstruction favoured by Hebbian
synapses have been studied in some of those works
\cite{GTA93,CT94,FKDO99,UHO04}. These are models with an underlying
asynchronous dynamics and phase diagrams were obtained which only
exhibit fixed-point solutions, in particular correlated attractors,
in accordance with a general expectation for networks with symmetric
interaction matrices far from the storage saturation limit in which
the ratio $\alpha=p/N$ of the number of stored patterns $p$ and the
number of neurons $N$ is zero, in the large $N$ limit. In contrast,
in the case of a synchronous dynamics with symmetric interactions
the stationary states may be either fixed-points or cycles of period
two, in the same limit \cite{LC95}. Rhythmic activity appears in
neurobiological systems \cite{Bu06} and the competition between
these features may yield interesting clues.

The presence of self-interactions of the units, which is consistent
with detailed balance in the synchronous dynamics of a network with
a symmetric interaction matrix, has not been considered so far
except in Little's model which has a simple Hebbian learning rule
\cite{Li74}-\cite{MT08}. The role of self-interactions which may be
either excitatory or inhibitory, is to control the fraction of spin
flips in the dynamics. Excitatory interactions may enhance the
retrieval performance while inhibitory interactions can give rise to
cyclic behaviour. Self-interactions and their relationship to
initial overlaps play a crucial role in Little's model leading to
frozen-in cycles of period two among other features, in the absence
of noise \cite{FK87,Fo88}. In a recent work it has been shown that
these cycles are destabilised in a slow dynamical process either by
synaptic or stochastic noise due to a macroscopic number of stored
patterns \cite{MT08}. This raises concern about the stability of
cycles of period two in general in the synchronous dynamics of
networks with symmetric interactions for the specific interesting
case of symmetric sequence processing competing with Hebbian
synapses.

There has been great interest in models with symmetric sequential
interactions due to the presence of correlated fixed-point
attractors \cite{CT94,FKDO99,UHO04,MAB03}, which are
stationary states that emerge from a balanced competition between
sequential and Hebbian synapses. They indicate a selectivity in
response to a set of previously learned uncorrelated patterns by
means of decreasing correlation coefficients for the attractors with
increasingly distant patterns from a stimulus. Correlated attractors
have been used to explain the results of experimental recordings of
a visual-memory task in the inferotemporal cortex of monkeys
\cite{MC88}-\cite{MAB03}. In the case of a synchronous dynamics, the
correlated fixed-point states might be destabilised by the presence
of a macroscopic number of flipping spins giving rise to oscillatory overlaps.

The asymptotic states of a feed-forward layered neural network model
for competing symmetric sequence processing with Hebbian synapses
have been discussed in a recent work \cite{MT07}. The model is
described by a synchronous dynamics and it is characterised by
asymmetric synaptic connections between units in two consecutive
layers and there are neither lateral synapses between units in the
same layer nor self-interactions of the units. Phase diagrams of
stationary states were obtained exhibiting retrieval states,
correlated states, symmetric mixture states and stable cycles of
period two, for increasingly larger fractions of sequential
synapses. All of these states are robust either to synaptic noise or
to stochastic noise due to a macroscopic number of stored patterns.

The purpose of the present paper is to study the synchronous
dynamics and the asymptotic states of a recurrent network model of
binary units and patterns for competing interaction between
symmetric sequence processing and Hebbian synapses, in order to
investigate the presence and stability of cycles of period two and
of other states which could be competing with the retrieval and
with the correlated states. We make use of a generating functional
approach (GFA) for the dynamics of disordered systems
\cite{AC01,DD78}, which is an exact procedure in the mean-field
limit, and we use an adaptation of the numerical simulation
procedure of Eissfeller and Opper (EO) \cite{EO92} based on the GFA
in order to implement the calculation of single-site averages. We
also resort to a recently introduced alternative approach
\cite{MT08}.

The outline of the paper is the following. We introduce the model in
section 2 and present a brief summary of the well known GFA and the
EO procedure in section 3 as well as explicit dynamic recursion
relations for the overlaps. We present our results for the phase
diagrams in section 4 and conclude with a summary and further
discussion in section 5.

\section{The model}

We consider a network of $N$ Ising neurons in a microscopic state
$\bsigma^t= ( \sigma_1^t,\dots,\sigma_N^t )$, at the discrete
time-step $t$ in which each $\sigma_i^t=\pm 1$ represents the state of
an active or inactive neuron, respectively. The states of all neurons are
updated simultaneously at each time-step according to the alignment
of each spin with its local field
\begin{equation}
h_i^t=\sum_{j}J_{ij}\sigma_j^t + \theta_i^t\,\,\,, \label{1}
\end{equation}
following a microscopic stochastic single spin-flip dynamics with
transition probability
\begin{equation}
w(\sigma_i^{t+1}|h_i^t) =\frac{1}{2}\big
[1+\sigma_i^{t+1}\tanh{(\beta h_i^t)}\big ]\,\,\, \label{2}
\end{equation}
ruled by the synaptic noise control parameter $\beta=T^{-1}$. Here,
$J_{ij}$ is the synaptic coupling between neurons $i$ and $j$ and
$\theta_i^t$ is an external stimulus. The dynamics is a
deterministic one when $T=0$ and fully random when $T =\infty$. In
the former case, $\sigma_i^{t+1}={\rm sgn}(h_i^t)$.

A macroscopic set $\bxi^{\mu} =(\xi_1^{\mu},\dots,\xi_N^{\mu})$,
$\mu=1,\dots,p$ of $p=\alpha N$ independent and identically
distributed quenched random patterns, each $\xi_i^{\mu}=\pm 1$ with
probability $\frac{1}{2}$, is embedded in the network by means of
the synaptic coupling $J_{ij}$ between distinct neurons $i$ and $j$.
One may think of $j$ and $i$ as pre- and post-synaptic neurons,
respectively, the activities of which give rise to that coupling. We
assume, as usual, that a finite number $c$ of patterns is condensed
so that the overlaps with the state of the network, defined below,
are finite and responsible for the signal in the local field. The
remaining macroscopic number of $p-c$ non-condensed patterns will
give rise to the noise in the local field. To be specific, we assume
that the condensed patterns are cyclic so that $\bxi^{c+1} =
\bxi^{1}$.

The non-condensed patterns need not be embedded in the network by
the same learning rule as that for the condensed patterns. Indeed,
one may think that those patterns have been learned in a previous
stage, in accordance with an argument that has been used before
\cite{CT94}. We make use of this freedom in order to simplify the
calculations by assuming a Hebbian rule for the non-condensed
patterns. Guided by work on the layered feed-forward network
\cite{MT07}, we expect qualitatively the same results as those
obtained here for the same learning rule for condensed and
non-condensed patterns. Thus, altogether, we take a synaptic
coupling
of the form\\
\begin{eqnarray}
  \fl
  J_{ij}&=\frac{\nu}{N}\sum_{\mu =1}^c
   \xi_i^{\mu} \xi_j^{\mu}+\frac{1-\nu}{N}\sum_{\mu =1}^c
   (\xi_i^{\mu} \xi_j^{\mu+1}+\xi_i^{\mu+1} \xi_j^{\mu})
+ \frac{1}{N}\sum_{\mu=c+1}^p\xi_i^{\mu}
  \xi_j^{\mu}\,\,\,\,\,\,\,    {\rm if} \,\,\,\,\, i\neq j \,\,\, \nonumber\\
  \fl &=J_0 \,\,\,\,\,\,\,\, {\rm if} \,\,\,\,\, i=j \,\,\,, \label{3}
\end{eqnarray}\\
in which each value of $\nu$ $(0\leq \nu \leq 1)$ defines a model so
that when $\nu=1$ we get Little's model with a Hebbian rule and when
$\nu=0$ we have the purely symmetric sequential model. The first and
the second summations are responsible for the signal in the local
field while the last summation is responsible for the noise and we
comment on that term in section 5. The self-interaction $J_0$ is a
real non-random variable which can take any positive or negative
value enhancing or inhibiting, respectively, the local field in the
form of a pattern-independent contribution $J_0\sigma_i^{t}$. It
either tends to enforce the actual state of unit $i$, if $J_0$ is
positive, or to switch the state if $J_0$ is negative.

\section{The dynamic generating functional approach}

The dynamical evolution of the system is described by the moment
generating functional \cite{AC01}
\begin{eqnarray}
\fl
Z(\bpsi) &= \Big \langle  \exp \Big ( - \rmi \sum_{i} \sum_{s=0}^{t}
\psi_{i}^s \, \sigma_{i}^s \Big )  \Big \rangle \nonumber \\
\fl &= \sum_{\bsigma^{0},\dots,\bsigma^{t}}
\mathrm{Prob}(\bsigma^{0},\dots,\bsigma^{t}) \exp \Big ( - \rmi
\sum_{i}
  \sum_{s=0}^{t} \psi_{i}^s \, \sigma_{i}^s \Big ) \,\,, \label{4}
\end{eqnarray}
where $\bpsi^s =(\psi_1^s
,\dots,\psi_N^s)$ is a set of auxiliary variables that serve to generate
averages of moments of the states and the brackets denote an average
over all possible paths of states with probability
\begin{equation}
\mathrm{Prob}(\bsigma^{0},\dots,\bsigma^{t})
 =p(\bsigma^{0}) \prod_{s=0}^{t-1} \prod_{i}
 \frac{\exp(\,\beta\,\sigma_{i}^{s+1} \,
h_{i}^{s})}{2\,\mathrm{cosh}(\beta\,h_{i}^{s})} \label{5}
\end{equation}
that follows from (2). Assuming that for $N\rightarrow \infty$
only the statistical properties of the stored patterns will
influence the macroscopic behaviour of the system, one obtains the
relevant quantities which are the overlap $m^{t}_{\mu}$ of $O(1)$
with any one of the condensed patterns $\bxi^{\mu}$, the two-time
correlation function $C_{t l}$ and the response function
$G_{t l}$, given by
\begin{equation}
m_{\mu}^{t} = \frac{1}{N} \sum_{i} \overline{\xi_{i}^{\mu} \langle
  \sigma_{i}^t \rangle}
  = \lim_{\bpsi \rightarrow 0} \frac{\rmi}{N} \sum_{i}
  \xi_{i}^{\mu}  \frac{\partial \overline{Z(\bpsi)}}{\partial
  \psi_{i}^{t}}\,\,,   \label{6}
\end{equation}
\begin{equation}
C_{t l} = \frac{1}{N} \sum_{i} \overline{\langle
  \sigma_{i}^{t} \sigma_{i}^{l} \rangle}\nonumber
=  -\lim_{\bpsi \rightarrow 0}
  \frac{1}{N} \sum_{i}  \frac{\partial^{2} \overline{Z(\bpsi)}}{\partial
  \psi_{i}^{l} \psi_{i}^{t} }\,\, \label{7}
\end{equation}
and
\begin{equation}
G_{t l} = \frac{1}{N} \sum_{i} \frac{\partial
\overline{\langle \sigma_{i}^{t} \rangle}}{\partial
  \theta_{i}^{l}} =  \rmi \, \lim_{\bpsi \rightarrow 0}
  \frac{1}{N} \sum_{i}  \frac{\partial^{2} \overline{Z(\bpsi)}}{\partial
  \theta_{i}^{l} \psi_{i}^{t}}\,\,\,\, (l < t)\,\,,  \label{8}
\end{equation}
where the bar denotes the configurational average with the
non-condensed patters $\{ \bxi^{\rho} \} $
($\rho = c+1, \dots ,p$) and the restriction $l < t$ is due to
causality. The two-consecutive-time correlation function has a
particular meaning since $q_{t}=C_{t , t-1}$ gives the fraction of
flipping spins between two consecutive times as $(1-q_{t})/2$.

Following the now standard procedure in which the disorder average
is done before the sum over the neuron states one obtains exactly,
in the large $N$ limit, the generating functional \cite{AC01,EO92}
\begin{eqnarray}
\fl \overline{Z(\bpsi)} = \Big{\langle}
\sum_{\bsigma^{0},\dots,\bsigma^{t}} p(\bsigma^{0})  \exp \Big
( -i\sum_i \sum_{s=0}^{t} \psi_i^s \, \sigma_i^s  \Big ) \nonumber \\
\prod_i \prod_{s < t} \Big{[} \int dh_i^s \, \delta(h_i^s
-h_{\mathrm{eff}}^s )\, \mathrm{w}( \sigma_i^{s+1} |h_i^s ) \Big{]}
\Big{\rangle}_{\{\phi_i^s \}} \label{9}
\end{eqnarray}
in which $p(\bsigma^0)= \prod_{i} p(\sigma_{i}^0)$ is the
probability of the initial microscopic configuration while $\langle
\dots \rangle_{\{\phi_{i}^{t} \}}$ denotes an average over a set of
temporarily correlated Gaussian random variables $\{\phi_{i}^{t}\}$ for unit
$i$, with zero-average and a correlation matrix given below. The
random variables on different units turn out to be uncorrelated and
one is left with a single-site effective theory in which a neuron
evolves in time according to the transition probability
\begin{equation}
 \mathrm{w}(\sigma^{t+1} | h_{\mathrm{eff}}^t )= \frac{1}{2} \Big [  1
+ \sigma^{t+1} \tanh{\big( \beta h_{\mathrm{eff}}^{t} \big) } \Big ]
\,\, \label{10}
\end{equation}
with an effective local field given by
\begin{equation}
  h_{\mathrm{eff}}^t = \sum_{\mu,\rho\leq c}\xi^{\mu}A_{\mu \rho}
   m^{t}_{\rho} + J_{0} \sigma^t
  + \alpha \sum_{s <t}   R_{t s}   \sigma^{s} +
  \sqrt{\alpha} \phi^t \,\,, \label{11}
\end{equation}
where
\begin{equation}
A_{\mu \rho}=\nu \delta_{\mu,\, \rho}+(1-\nu)\,(\delta_{\mu,\,
\rho+1} + \delta_{\mu,\, \rho-1}) \,\,,  \label{12}
\end{equation}
and we assumed that $\theta^t=0$. The two non-trivial contributions
to the effective local field for $\alpha>0$ come from a retarded
self-interaction involving the matrix elements
\begin{equation}
R_{t s} = [\bG(\bI-\bG)^{-1}]_{t s} \label{13}
\end{equation}
and the zero-average temporarily correlated Gaussian noise $\phi^t$
with variance
\begin{equation}
S_{t s} = \langle \phi^t \phi^s \rangle_{\{\phi^t \}} =
[(\bI-\bG)^{-1}\bC(\bI-\bG^{\dagger})^{-1}]_{t s}\,\,.\\
\label{14}
\end{equation}
Here, $\bC$ and $\bG$ are matrices with elements $\{ C_{t s} \} $
and $\{ G_{t s} \}$, respectively. Both contributions account for
memory effects in the network that come from the noise in the
original local field due to the macroscopically large number of
non-condensed patterns.

The dynamics of each of the macroscopic quantities, given
by (\ref{6}-\ref{8}), is obtained from the statistics
of the effective single neuron process through the average
\begin{equation}
\langle f(\bsigma) \rangle_{*} = \int d\bphi \mathrm{P}(\bphi)
  \sum_{\bsigma}  \mathrm{P}(\bsigma|\bphi) f(\bsigma) \,\,, \label{15}
\end{equation}
where $\bsigma=\{\sigma^t \}$ and $\bphi=\{\phi^t \}$ are now
single-site vectors that follow a path in discrete times, and
\begin{equation}
 \mathrm{P}(\bsigma|\bphi)=p(\sigma^0) \prod_{s<t}
 \mathrm{w}(\sigma^{s+1} | h_{\mathrm{eff}}^s ) \,\, \label{16} \\
\end{equation}
is the single-spin path probability given the Gaussian noise $\bphi$
in the effective field, with a distribution
\begin{equation}
\mathrm{P}(\bphi)=\frac{1}{\sqrt{(2\pi)^t(\det\bS)}}
\exp \Big( {-\frac{1}{2} \bphi. \bS^{-1}\bphi} \Big)\,\,. \label{17}
\end{equation}

In order to obtain the full dynamic description of the transients
for $\alpha > 0$, we make use of the procedure of Eissfeller
and Opper in which the effective single-site dynamics given by
(\ref{10}-\ref{17}) is simulated by a Monte-Carlo method. There
are no finite-size effects, but a large number $N_{T}$ of stochastic
trajectories has to be generated for the single-site process in order
to keep the numerical error small. The macroscopic parameters can
then be obtained from the average
\begin{equation}
\langle f(\bsigma) \rangle_{*} = \frac{1}{N_{T}} \sum_{a=1}^{N_{T}}
f(\bsigma_{a})  \,\,, \label{18}
\end{equation}
where $\bsigma_{a}$ denotes the spin along the path $a$. The number
of stochastic trajectories $N_{T}$ should not be confused with the
number of neurons $N$, which goes to infinity. The specific algorithm
that implements the EO method is described in the literature \cite{EO92,Ve03}.

For a finite loading of patterns ($\alpha=0$) we can obtain
analytically recursion relations for the condensed overlaps and
expressions for the correlation coefficients defined below. In this
case the effective local field (\ref{11}) assumes the form
\begin{equation}
h_{\mathrm{eff}}^t = \sum_{\mu,\rho\leq c}\xi^{\mu}A_{\mu \rho}
   m^{t}_{\rho} + J_{0} \sigma^t \,\,, \label{19}
\end{equation}
so that $\mathrm{P}(\bsigma|\bphi)$ becomes $\bphi$ independent and
the integral over $\bphi$ in (\ref{15}) equals unity. Now
$h_{\mathrm{eff}}^t$ still depends on the microscopic state of the
system at time step $t$ and in order to calculate the sum over the
paths in the average $\langle \sigma^t \rangle_{*}$ we follow the
procedure introduced in \cite{MT08}. Assuming an initial
distribution $p(\sigma^0 ) = \frac{1}{2} [1 + \sigma^0 \xi^{\lambda}
m_{\lambda}^0 ]$, which corresponds to an initial vector overlap
with components $m_{\mu}^0 = \delta_{\mu \lambda} m_{\lambda}^0$
($\mu = 1,\dots,c$), the following system of recurrence relations
can be derived
\begin{eqnarray}
\fl \langle \sigma^{t+1} \rangle_{*} = \frac{1}{2} ( 1 + \langle
\sigma^t \rangle_{*} ) \tanh{\beta(\bxi. \bA \bm^t + J_{0})} +
\frac{1}{2} ( 1 - \langle \sigma^t \rangle_{*} )
\tanh{\beta(\bxi. \bA \bm^t - J_{0})} \,\,, \label{20} \\
\fl m^{t}_{\mu} = \Big{\langle} \xi^{\mu} \langle \sigma^t \rangle_{*}
\Big{\rangle}_{\bxi} \,\,, \label{21}
\end{eqnarray}
with $\bxi. \bA \bm^t =\sum_{\mu,\rho\leq c}\xi^{\mu}A_{\mu \rho}
m_{\rho}^t$ and $\langle \sigma^0 \rangle_{*} = \sum_{\sigma^0 }
p(\sigma^0) \sigma^0 = \xi^{\lambda} m_{\lambda}^0$. At a time step
$t$, one has to update all the $2^c$ possible values of
the single-site average $\langle \sigma^{t} \rangle_{*}$ by means of
(\ref{20}), each one related to a given realization of $\bxi$. This
procedure allows to calculate the overlaps at the same time step through
(\ref{21}). For $\nu=1$ and $c=1$, (\ref{20}) and
(\ref{21}) can be written as a single recurrence relation for the
overlap with one condensed pattern and one recovers the results
for Little's model \cite{MT08}.

In this work we are also interested in studying the correlation
between the stationary states of the network generated by different
initially stimulated patterns. Defining $\{ \langle
\sigma_{i}^{\lambda} \rangle \} $ ($i = 1, \dots, N$) as the
stationary state corresponding to a stimulus in pattern $\lambda$,
represented by an initial condition on the overlaps of the form
$m_{\mu}^0 = \delta_{\mu \lambda} m_{\lambda}^0$ ($\mu =
1,\dots,c$), the normalised correlation coefficient between two
stationary states is defined by \cite{CT94}
\begin{equation}
C_{\lambda \rho} = \frac{\sum_{i=1}^{N} \langle \sigma_{i}^{\lambda}
\rangle \langle \sigma_{i}^{\rho} \rangle} {\sum_{i=1}^{N} \langle
\sigma_{i}^{\lambda} \rangle^2 }\,\,. \label{22}
\end{equation}
We may use the self-average property to write (\ref{22}) in the
large-N limit, for $\alpha=0$, as
\begin{equation}
C_{\lambda \rho} = \frac{\big \langle \langle \sigma^{\lambda}
\rangle_{*} \langle \sigma^{\rho} \rangle_{*} \big \rangle_{\bxi} }
{\big \langle \langle \sigma^{\lambda} \rangle_{*}^2 \big
\rangle_{\bxi}}\,\,, \label{23}
\end{equation}
where $\langle \sigma^{\lambda} \rangle_{*}$ and $\langle
\sigma^{\rho} \rangle_{*}$ are determined by the fixed-point
solutions of (\ref{20}) and (\ref{21}). When $\alpha\neq 0$ the
summations over sites in (\ref{22}) can no longer be replaced by the
averages over patterns in (\ref{23}), but one may use the similarity of
the overlap vector with that at $\alpha=0$ as a guide to decide if
one is in the presence of a correlated state or not. In this model,
the structure of the stationary overlap vector is the
same when different initially stimulated patterns are considered.
Thus the correlation coefficient $C_{\lambda \rho}$ depends only on
the distance $d=|\lambda-\rho|$ between the patterns in the
sequence. The correlated stationary states one is interested in a
visual-memory task are those for which the decreasing correlation coefficients
vanish (or almost vanish) for increasing $d$, indicating a clear
selectivity with respect to the patterns in the sequence.

\section{Results}

We focus mainly on phases of retrieval, cyclic and correlated
fixed-point states. All the explicit results shown in this section
were obtained assuming an initial overlap $m^{0}_{\mu} =
m^{0}_1\delta_{\mu 1}$ ($\mu=1,\dots,c$), with $m^{0}_1=0.4$. The
reason for this choice in place of the more
\begin{figure}[ht]
\center
\includegraphics[scale=0.48]{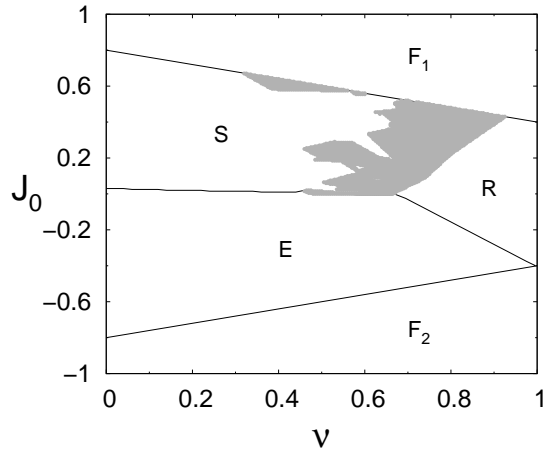}
\caption{Phase diagram of stationary states for $c=10$ condensed
patterns, $\alpha=0=T$ and initial overlap $m^{0}_{\mu} = 0.4
\delta_{\mu 1}$ ($\mu = 1,\dots,c$). $R$ is a retrieval phase, $S$
is a phase of symmetric-like states, $E$ is a phase of period-two
cycles, the grey area is a phase of correlated fixed-point states,
$F_1$ and $F_2$ are phases of frozen-in fixed-point and period-two
cyclic states, respectively.}
\label{fig1}
\end{figure}
popular $m^{0}_1=1$ which favours the retrieval phase is to be
within the basin of attraction of the other phases of interest for
convenient values of $J_0$.

We consider in this work $c=10$ condensed patterns in all the cases
studied, which is suitable due to the following. First, an
interesting sequence for associative-memory tasks should not be too
short. Second, we expect that already for a value of $c$ of this
size the phase diagrams should only exhibit small quantitative
differences in the phase boundaries for different values of $c$,
guided by the results for the feed-forward network \cite{MT07}.
Third, we are interested in fixed-point correlated states
\cite{GTA93,CT94}, characterised by well-defined correlation
coefficients $C_{d}$ that decrease down to a vanishingly small value
for the largest $d$ which should not be too small.

Although only few results can be derived analytically (see below)
due to the complexity of the problem, all the features of the phase
diagrams can be obtained numerically for $\alpha=0$. We show first
the results in that case for the stationary behaviour obtained by
means of the iteration of (\ref{20}) and (\ref{21}) until a
stationary overlap vector $\bm$ is reached. The $(\nu,J_{0})$ phase
diagram of stationary states in the absence of noise ($\alpha=0=T$)
is shown in figure \ref{fig1}. Given a value of $\nu$ (which defines
a model) and the initial overlap with the condensed patterns that
specifies the basins of attraction of the phases of interest, the
size and the sign of the self-interaction $J_0$ yields the various
stationary phases as indicated. The variety of the phases is also
investigated studying the behaviour of the system as a function of
the model parameter $\nu$, for a given $J_0$.

For large values of $|J_0|$, there is a phase $F_1$ of frozen-in
fixed-points for positive $J_0$, with an overlap $m_{\mu} =0.4\,
\delta_{\mu 1}$ ($\mu = 1,\dots,c$) that stays the same as the
initial overlap for all times, and there is a phase $F_2$ of
frozen-in cycles for negative $J_0$, with an overlap $m_{\mu}^{t} =
(-1)^t 0.4\, \delta_{\mu 1}$ ($\mu = 1,\dots,c$) that keeps
switching between the initial overlap and its opposite. The phase
boundaries of $F_1$ and $F_2$ can be derived analytically from
(\ref{20}) and (\ref{21}) in the $T\rightarrow 0$ limit iterating
the condensed overlaps at consecutive times. Writing
the initial overlap as $m^{0}_{\mu} =  m^{0}_{\lambda} \delta_{\mu \lambda}$
($\mu=1,\dots,c$), for a general $m^{0}_{\lambda}$ $(0\leq m^{0}_{\lambda}
\leq 1)$, this yields first an expression for $m^{1}_{\lambda}$ at the first time
step in terms of $J_0$, $\nu$ and $m^{0}_{\lambda}$. The conditions
that $m^{1}_{\lambda}= m^{0}_{\lambda}$ or $m^{1}_{\lambda}=
-m^{0}_{\lambda}$ (the case for $F_1$ or $F_2$, respectively) are
\begin{eqnarray}
\fl J_0>m^{0}_{\lambda} (2-\nu)\,\,\,,\,\,\, \rm phase\,\, F_1 \label{24}\\
\fl J_0<m^{0}_{\lambda} (\nu-2)\,\,\,,\,\,\, \rm phase\,\, F_2 \label{25} \,\,\,,
\end{eqnarray}
which are independent of $c$. These conditions also lead to
$m^{1}_{\lambda\pm n}=0$ ($n=1,\dots,c-1$) which ensures that
the network state does not switch from one condensed pattern
to another. The same
relation $m^{1}_{\mu} = \pm m^{0}_{\lambda} \delta_{\mu \lambda}$
($\mu=1,\dots,c$) holds from any time-step to
the next one giving rise to the phase boundaries of the frozen-in
states. In phases $F_1$ and $F_2$ the system is not useful for
information processing but, as will be seen below, the frozen-in
states become destabilised in the presence of synaptic noise ($T>
0$) and, eventually, lead to dynamically useful fixed-point or
oscillating states.

In figure \ref{fig1} there is a phase $R$ of retrieval fixed-point
states for large $\nu$ that reflects the dominance of the Hebbian
synapses, with stationary overlaps $m_{\mu} = \delta_{\mu 1}$ ($\mu
= 1,\dots,c$). The upper and lower phase boundaries of $R$ end at
$J_0=\pm 0.4$ for $\nu=1$ which has been obtained analytically
before for Little's model in the absence of noise \cite{Fo88,MT08}.
A phase $S$ of symmetric or symmetric-like states of equal or
similar overlap components, respectively, appears for
not too large $J_0 \geq 0$. This phase exhibits a succession of
multiple discontinuous transitions of the overlap vector $\bm$ for
intermediate values of $\nu$ that will be shown below in connection
with the phase of correlated states, which is the grey area in the
figure \ref{fig1}. It is appropriate to note here that the latter is
a phase that arises from the competition between sequential and
Hebbian processing and it is not present in Little's model with pure
Hebbian synapses. There is also a phase $E$ of period-two cycles
with $m_{\mu}^{t+2}=m_{\mu}^{t}$ for $\mu=1,\dots,c$ mostly for
negative $J_0$, as shown below in figure $2$ for one of the overlap
components. Phase $E$ exhibits a similar behaviour as that in phase
$S$ with multiple transitions but now to a variety of period-two
cycles, in place of fixed-point states.

\begin{figure}[ht]
\center
\subfigure[$\,J_0 =-0.1$.]{
\includegraphics[scale=0.48]
{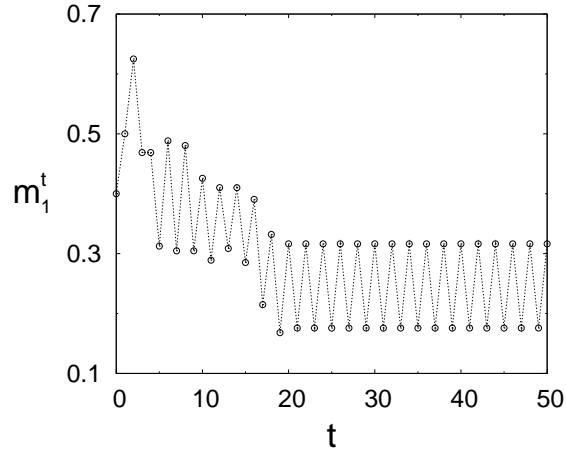} \label{fig2a}}
\subfigure[$\,J_0=-0.3$.]{
\includegraphics[scale=0.48]
{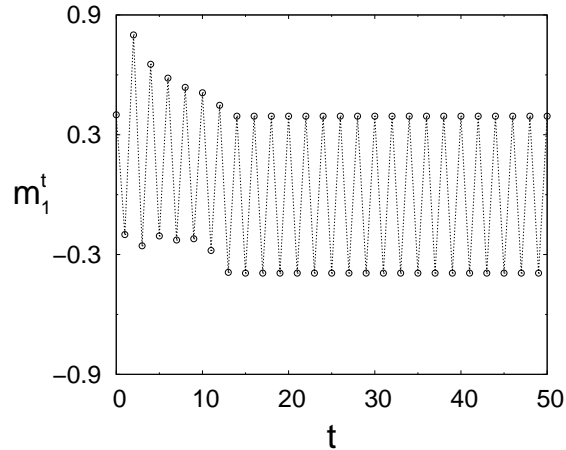} \label{fig2b}}
\caption{Evolution of a single overlap (the
other components behave in a similar way) in phase $E$ for
$\nu=0.3$, initial overlap $m^{0}_{\mu} = 0.4\delta_{\mu 1}$ and
$\alpha=0=T$ in (a) the upper part of the phase with $J_0=-0.1$, and
in (b) the lower part with $J_0=-0.3$.}
\label{fig2}
\end{figure}

In figure \ref{fig2} we show the period-two cyclic behaviour within
phase $E$ for a single component, the other components behave in a
similar way, for a typical $\nu=0.3$ and two negative values of
$J_0$ within that phase, as indicated. In the upper part of the
phase the stationary oscillation is between positive overlaps and in
the lower part the oscillation is between $\pm m$.
\begin{figure}[ht]
\center
\subfigure[$\,J_0 =0.2$.]{
\includegraphics[scale=0.48]
{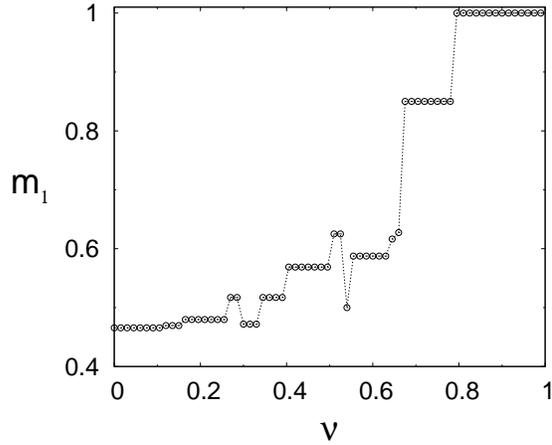} \label{fig3a}}
\subfigure[$\,\nu=0.6$.]{
\includegraphics[scale=0.48]
{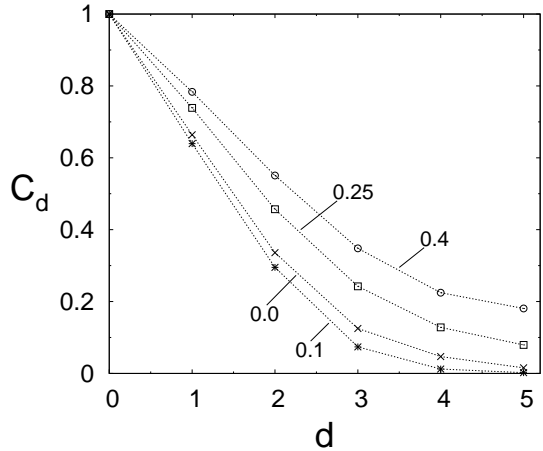} \label{fig3b}} \caption{Multiple transitions of one of
the overlap components within the whole range of $\nu$ (a) and
correlation coefficients $C_d$ defined in the text for various
values of $J_0$ (b). Both figures were generated for $c=10$,
$\alpha=0=T$ and initial overlap $m^{0}_{\mu} = 0.4 \delta_{\mu 1}$
($\mu = 1,\dots,c$).} \label{fig3}
\end{figure}

The fixed-point solutions for one of the overlap components at
$\alpha=0=T$ for a typical $J_0=0.2$ in phase {\bf $S$} and the
whole range of $\nu$ is shown in figure \ref{fig3a}. There is a
finite number of bifurcations at specific values of $\nu$, in both
the white and gray regions in that phase ending at the retrieval
phase with $m_1=1$. All the other overlap components follow a
similar behaviour with transitions not necessarily of the same size
but at the same values of $\nu$. We have studied the fixed-point
solutions in both symmetric and correlated states calculating the
correlation coefficients $C_d$ as a function of the distance $d$, as
shown in figure \ref{fig3b} for a fixed $\nu=0.6$ and different
values of $J_0$, as indicated. Either $C_d$ decreases to a finite
value, which is typical of symmetric-like fixed-points, or $C_d$
decreases to zero, which is a characteristic of correlated states.
The numerical criterion chosen for the latter employed in the
construction of the gray region of figure \ref{fig1} is that
$C_5<0.02$ for the maximum distance $d=5$. The non-monotonic
behaviour of $C_d$ with $J_0$ reflects the reentrance region to the
phase of correlated states in figure \ref{fig1}.
\begin{figure}[ht]
\center
\includegraphics[scale=0.48]{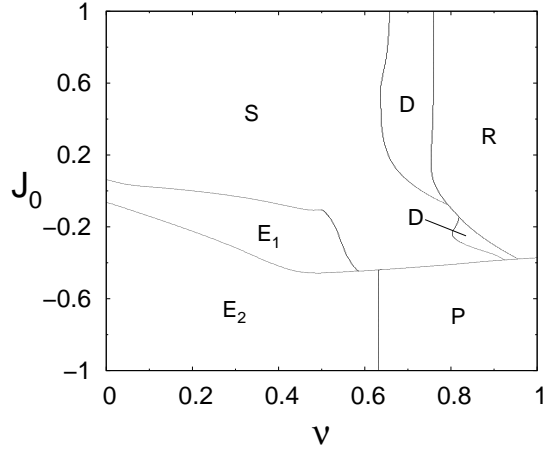}
\caption{Phase diagram of stationary states for $c=10$, $\alpha=0$,
$T=0.2$ and initial overlap $m^{0}_{\mu} = 0.4\, \delta_{\mu 1}$ ($\mu
= 1,\dots,c$). $R$ is a retrieval phase, $S$ is a phase of
symmetric-like states, $P$ is a paramagnetic phase, the regions $D$
are phases of correlated fixed-point states and $E_1$ and $E_2$ are phases
of cyclic states.}
\label{fig4}
\end{figure}

In order to illustrate the role of synaptic noise on the behaviour
of the network, we show in figure \ref{fig4} the $(\nu,J_0)$ phase
diagram of stationary states for $\alpha=0$ and $T=0.2$. For $\nu \approx
1$, the model has a similar behaviour to that of Little's model
\cite{Fo88,MT08}. There is a retrieval phase $R$ with $\bm \simeq
(m_{1},0,\dots,0)$ and $m_{1}$ assuming values in the range $0.9
\lesssim m_{1} \lesssim 1$, depending on the parameters $\nu$ and
$J_0$. The frozen-in cyclic states in the phase $F_2$ at
$T=0=\alpha$ become destabilised by synaptic noise for all $\nu$.
For the larger $\nu$ they go into a paramagnetic phase with $\bm=0$
and for the smaller $\nu$ they become period-two cycles with
overlaps that evolved from the initial value. The cyclic states in
region $E_1$ are reminiscent of those in the feed-forward network
\cite{MT07}, with each overlap component oscillating between a
larger and a smaller positive value. In fact, for $\alpha=0=J_0$ we
recover the results of \cite{MT07}, since in this case the equations
for the order parameters are precisely the same in the layered and
in the recurrent networks. The overlap components in phase $E_2$ are
$m^{t}_{\mu} =(-1)^{t} m_{\mu}$, each one exhibiting usually a
different amplitude {\bf $m_{\mu}$}. The frozen-in states
in phase $F_1$ for $\alpha=0=T$ are destabilised by synaptic noise and
become symmetric or symmetric-like states in phase $S$. This is now a
phase that ends at two phases of correlated fixed-point solutions in
the two disjoint regions $D$, that differ in the rate at which $C_d$
goes to zero.
\begin{figure}[ht]
\center
\includegraphics[scale=0.48]{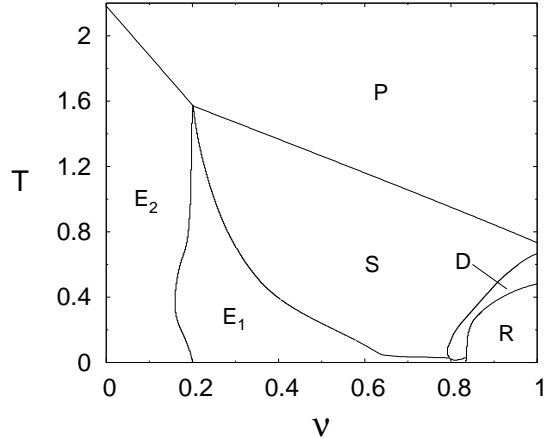}
\caption{Phase diagram of stationary states for $c=10$, $\alpha=0$,
$J_0=-0.2$ and initial overlap $m^{0}_{\mu} = 0.4 \delta_{\mu 1}$
($\mu = 1,\dots,c$). The phases are those of figure \ref{fig4}.}
\label{fig5}
\end{figure}
In the lower region $D$, we have $C_d \approx 0$ for $d=3$ whereas
in the upper region $D$, $C_d \approx 0$ for $d=5$. As can be seen
from figure \ref{fig4}, the range of values of $\nu$ where the
network evolves to correlated fixed-point states can be enhanced by
an increase of $J_0$.

To illustrate the robustness of
the different phases with respect to synaptic noise, we
show in figure \ref{fig5} the $(\nu,T)$ phase diagram for $\alpha=0$
and a small $J_0=-0.2$.
\begin{figure}[ht]
\center
\subfigure[]{
\includegraphics[scale=0.48]
{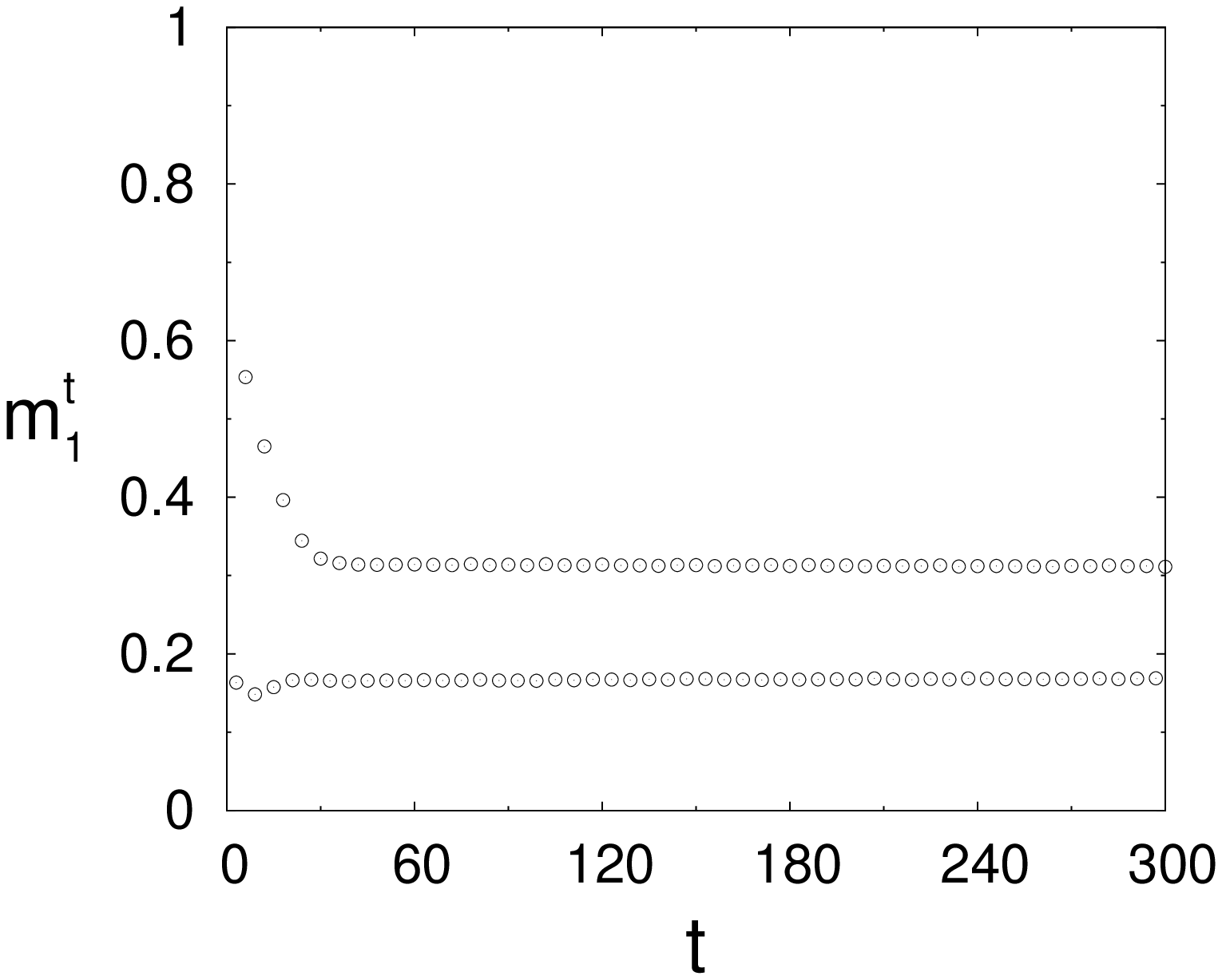}
\label{fig6a} }
\subfigure[]{
\includegraphics[scale=0.48]
{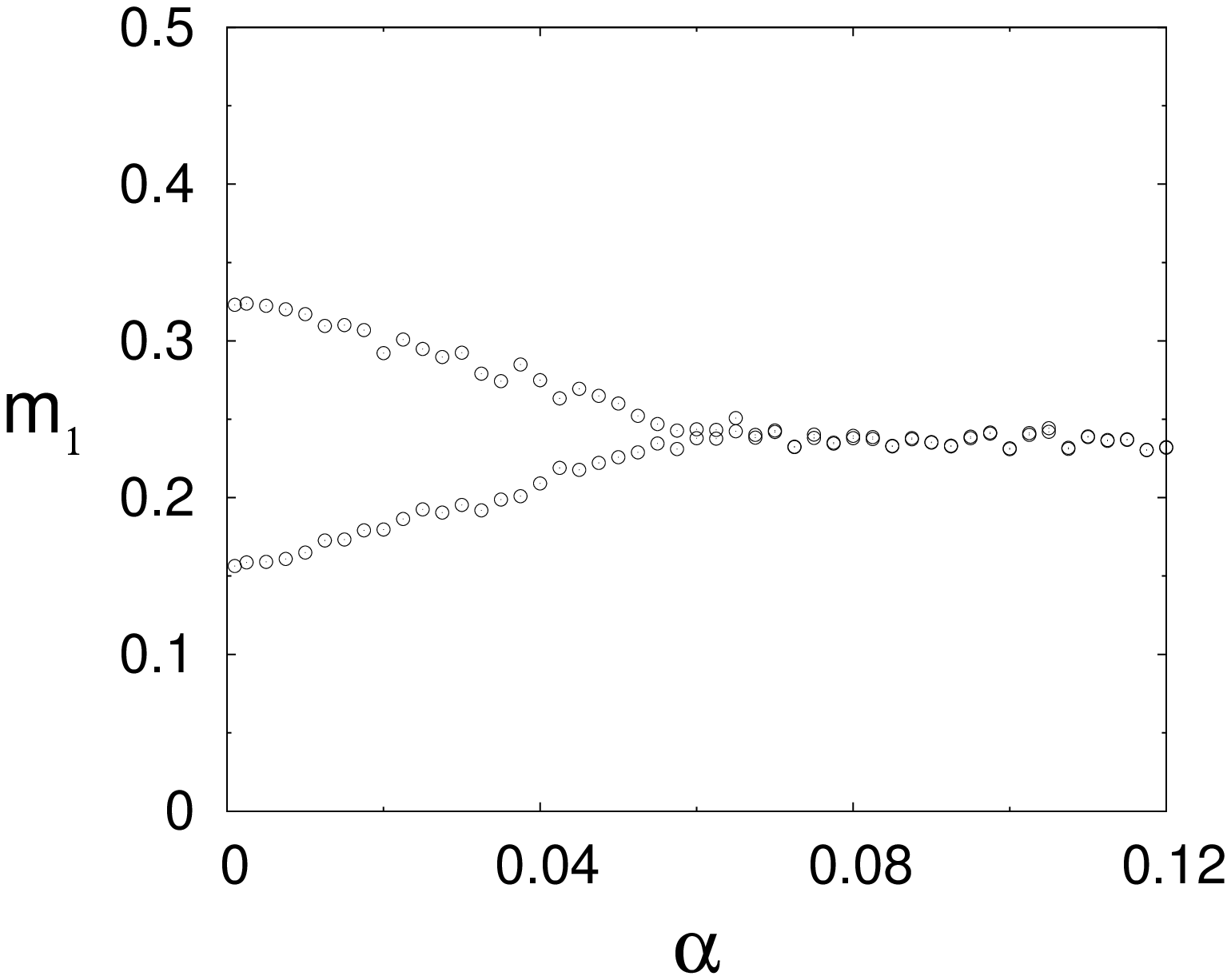}
\label{fig6b} }
\caption{Dynamics of one overlap
component $m^{t}_{1}$ in the cyclic phase $E_1$ for $\alpha=0.01$ (a),
and stationary overlap for increasing stochastic noise $\alpha$ with
transition to the symmetric-like phase $S$ at $\alpha\approx 0.06$ (b),
both for $c=10$, $T=0.2$, $J_0=-0.02$, $\nu=0.1$ and initial overlap $m^{0}_{\mu} = 0.4
\delta_{\mu 1}$ ($\mu = 1,\dots,c$). These results were generated by
the Eissfeller-Opper procedure with $N_T=5 \times 10^5$ trajectories.}
\label{fig6}
\end{figure}
Although the oscillation amplitudes of the overlap components in phases
$E_1$ and $E_2$ decrease with increasing $T$, the cyclic solutions are
stable even for a relatively large synaptic noise.

We consider now the effects of stochastic noise due to a macroscopic
number of patterns, $p=\alpha N$, employing the procedure of
Eissfeller and Opper \cite{EO92}.
\begin{figure}[ht]
\center
\includegraphics[scale=0.48]
{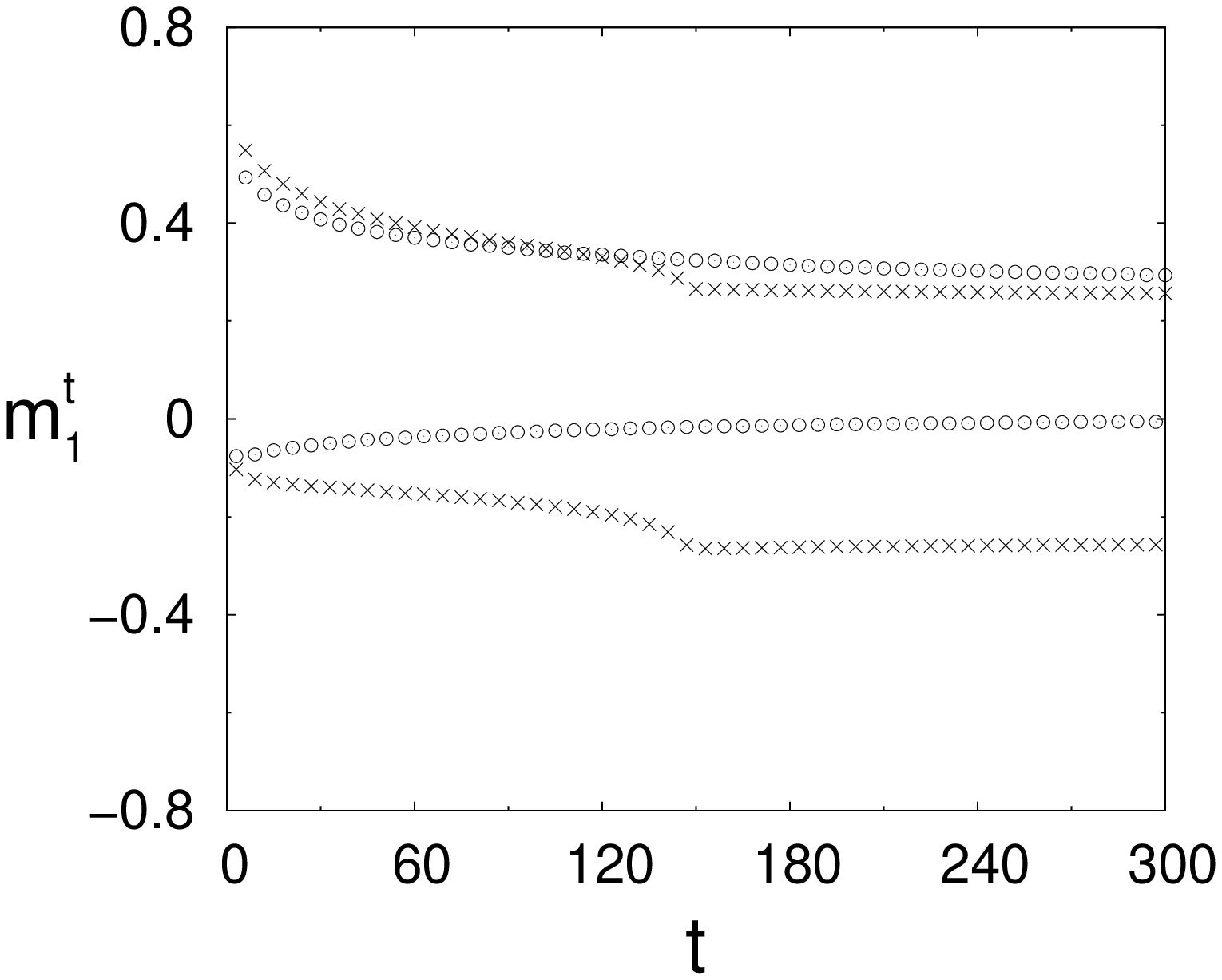}
\caption{Dynamics
of $m^{t}_{1}$ in the cyclic phase $E_2$ for $c=10$, $J_0=-0.3$,
$\nu=0.1$,
 $T=0.2$, initial overlap $m^{0}_{\mu} = 0.4
\delta_{\mu 1}$ ($\mu = 1,\dots,c$) and two levels of stochastic
noise: $\alpha=0.5$ (crosses) and $\alpha=0.7$ (circles). These
results were generated by the Eissfeller-Opper procedure with $N_T=5
\times 10^5$ trajectories. } \label{fig7}
\end{figure}
Since the construction of a phase diagram using this method is a
prohibitive task due to the slow dynamics for finite $\alpha$
\cite{MT08}, we concentrate on the stability of some typical states. First
we study the cyclic states favoured by dominating
sequential synapses, that is for small $\nu$. In figures \ref{fig6a}
and \ref{fig6b} we illustrate, respectively, the dynamics of $m_{1}^{t}$
and the stability to stochastic noise of a stationary overlap component (the
other components behave in a similar way), both for a state
in phase $E_1$, when $J_0=-0.02$,
$\nu=0.1$ and $T=0.2$. Figure \ref{fig6a} shows, for $\alpha=0.01$,
that $m_{1}^{t}$ keeps oscillating between the upper and lower values
at consecutive time-steps without any significant variation
in the amplitude with the asymptotic state being reached for $t \sim
40$ time steps, suggesting that the cyclic states in phase $E_1$ are
stationary states of the network dynamics for small values of
$\alpha$. The cycles decrease in amplitude with increasing $\alpha$
within that phase and change into symmetric-like states for larger
$\alpha$, as shown in figure \ref{fig6b}.

The cyclic states in phase $E_2$ turn out to be stable for higher
stochastic noise as shown in figure \ref{fig7} by the dynamics of
$m_{1}^{t}$ for $J_0=-0.3$, $\nu=0.1$, $T=0.2$ and two values of
$\alpha$. Indeed, for $\alpha=0.5$ the overlap component keeps
oscillating with no change in the amplitude after a transient period,
indicating stability to stochastic noise, whereas for $\alpha=0.7$ the
amplitude is already decreasing, indicating that the cycles are unstable for that
load of patterns. The reason for the increased robustness to stochastic
noise of the cycles in phase $E_2$, in contrast to those in phase $E_1$, is
that the former are deeper in the cyclic region, with a larger
negative $J_0$ for the same initial overlap $m_{\mu}^0$. We comment in
the last section on the relative robustness to stochastic noise of both
cyclic phases.

\begin{figure}[ht]
\center
\includegraphics[scale=0.48]
{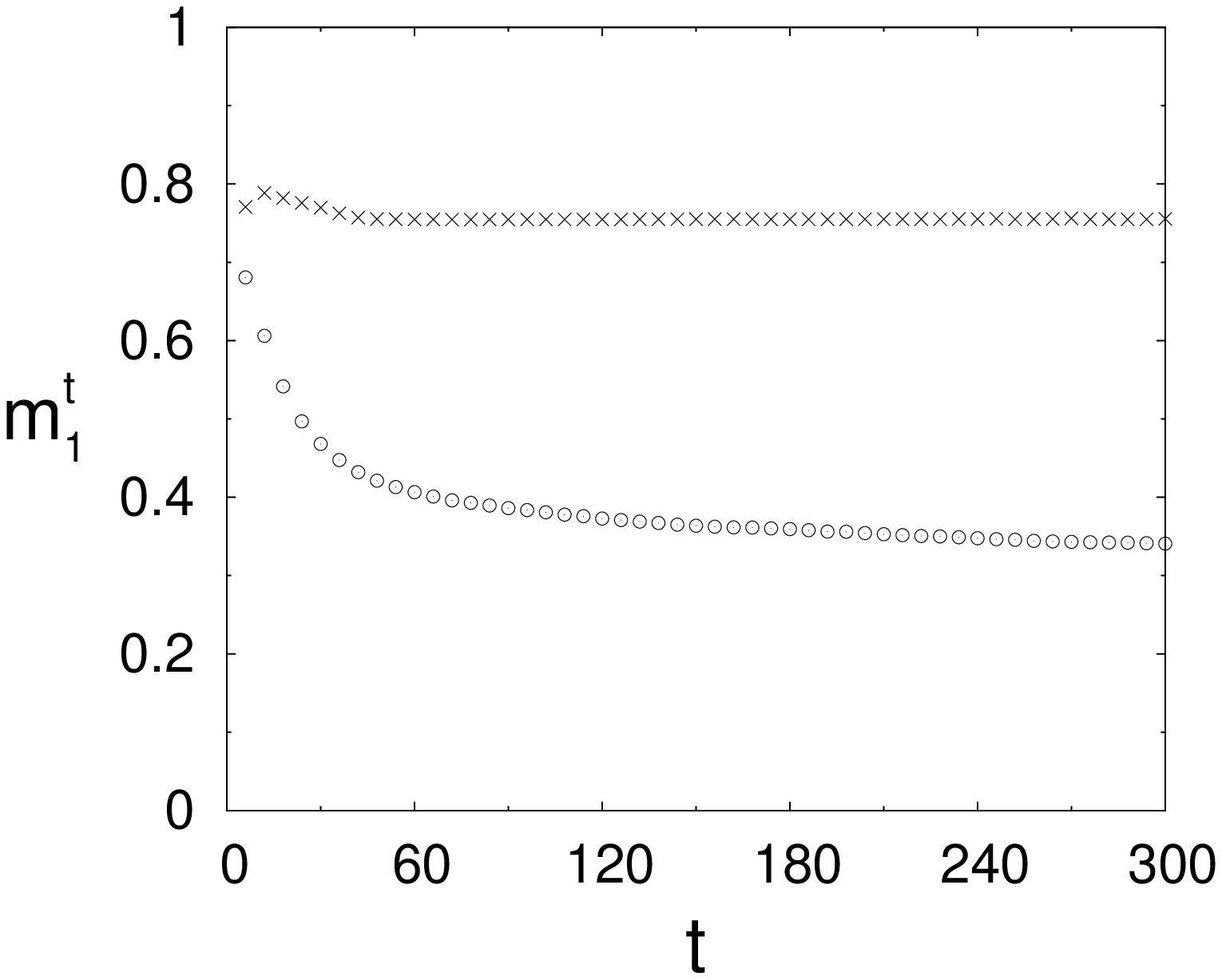} \caption{Dynamics of $m^{t}_{1}$ in the lower phase $D$
of correlated fixed-point states for $c=10$, $J_0=-0.25$,
$\nu=0.83$, $T=0.005$, initial overlap $m^{0}_{\mu} = 0.4
\delta_{\mu 1}$ ($\mu = 1,\dots,c$) and two levels of stochastic
noise: $\alpha=0.006$ (crosses) and $\alpha=0.01$ (circles). These
results were generated by the Eissfeller-Opper procedure with $N_T=5
\times 10^5$ trajectories. } \label{fig8}
\end{figure}

We consider next the stability of correlated fixed-point states in
the presence of stochastic noise which are expected for dominating
Hebbian synapses in the presence of sequential interactions and
we resort again to the EO procedure.
The dynamics of $m^{t}_{1}$ up to $t=300$ time steps was studied for
a state
within each region $D$ of figure \ref{fig4}. In figure \ref{fig8}
are shown results in the
lower region $D$ with $\nu = 0.83$, $J_0 =-0.25$,
and $T=0.005$ in order to extract mainly the effects of
stochastic noise for two values of $\alpha$. The upper curve, for
$\alpha=0.006$, indicates that the correlated fixed-point
state is stable with a stationary overlap vector given
by $\bm \simeq (0.75,0.25,0, \dots,0,0.25)$. This state is
already unstable for a somewhat larger $\alpha=0.01$, as suggested
by the lower curve, since the overlap $m^{t}_{1}$ is decreasing
towards a value that is quite different from $m_{1} \simeq 0.75$. In
fact, we obtained an overlap vector given approximately by $\bm \simeq
(0.34,0.30,0.21,0.14,0.09,0.07,0.09,0.14,0.20,0.28)$, indicating
that the network is evolving to a symmetric-like state.

We have also investigated the stability of states in the upper
region $D$ and in the retrieval phase, and found similar results to
those in the lower $D$ region, in the first case, and results
reminiscent to those for Little's model, in the second case,
indicating stability for small values of $\alpha$.

\section{Summary and conclusions}

The generating functional approach has been used in this work to
study the synchronous dynamics, the stationary states and the
transients of a recurrent neural network model with synapses
generated by the competition between symmetric sequence processing
and Hebbian pattern reconstruction. Either the numerical procedure
of Eissfeller and Opper, based on the GFA, to simulate paths of
single-spin states or a simpler alternative procedure have been used
in this work to obtain results in the presence or absence of
stochastic noise due to the load of a macroscopic number of
patterns. There is a single time scale in the dynamics (the step of
unit size) leading to both fixed-point and cyclic behaviour of
period two. The latter arises from the synchronous updating of all
units at every time step and it is enhanced by two features: the
sequential interactions and the self-interaction of the units. The
mean-field dynamics done here allows us to study the stability of
the fixed-point and cyclic states as well as the transitions between
them.

In distinction to Little's model (the case $\nu=1$ of purely Hebbian
synapses) where the cycles of period two only appear as frozen-in
states in the absence of noise and become destabilised by synaptic or
stochastic noise, there appear now in the case where $0\leq\nu<1$ also stable
dynamic cycles of period two, either with or even without noise. These are
cycles that evolve from the initial overlap to a stationary state, and they
appear in a large region of the phase diagram.

The retrieval behaviour and the fixed-point correlated states are
also enhanced by the presence of a self-interaction and in this work
we investigated the changes in the phase diagrams due to that
interaction. Phase diagrams of stationary states were obtained in
this work and it was shown that fixed-point correlated
states are clearly separated from both phases of cyclic states
already for a small but finite synaptic noise, independently of the
size (and even in the absence) of a self-interaction. This suggests,
within the limited conclusions that can be drawn from an attractor
neural network model, that there should be no interference of the
oscillating states produced by the synchronous dynamics with the
correlated fixed-point states that are crucial in visual task
experiments.

We comment, next, on the last summation in the synaptic interaction
given by (\ref{3}), which is responsible for the noise term in the
local field $h_i^{t}$. One may consider a more general form
$\frac{1}{N}\sum_{\mu,\,\rho>c}^p\xi_i^{\mu}B_{\mu\,\rho}
  \xi_j^{\rho}$\,
with an interaction matrix of the same form as (\ref{12}) for the
condensed part,
\begin{equation}
B_{\mu \rho}=b \delta_{\mu,\, \rho}+(1-b)\,(\delta_{\mu,\, \rho+1} +
\delta_{\mu,\, \rho-1}) \,\,
\end{equation}
with $0\leq b \leq 1$, which could be equal to $\nu$. The case we
considered here, for simplicity, is a Hebbian noise with $b=1$. The
more general form has been used in the case of the layered
feed-forward network \cite{MT07} and one may infer from the results
of that work the qualitative changes on the results presented here
when $b=\nu$. It turns out that the pure Hebbian case underestimates
slightly the storage capacity of the fixed-point states for large
values of $\nu$. On the other hand, the storage capacity for almost
pure cyclic behaviour, with small $\nu$, is overestimated by a pure
Hebbian noise, and this is one of the reasons for the large value of
$\alpha$ for which the cyclic states are still stable in both cyclic
phases $E_1$ and $E_2$, as found in section 4.

Finally, there are a few features of the model which are worth
pointing out. First, is that the results obtained with the symmetric
interactions $J_{ij}$ in (\ref{3}), are quite different from those
for Little's model. One of these results is the presence of
fixed-point correlated states, another one is the presence of stable
cycles in phases $E_1$ and $E_2$, in distinction to the absence of
cycles in Little's model with noise. Furthermore, excitatory
self-interactions enhance fixed-point correlated states as shown by
the enlarged upper part of the $D$ phase. Also, inhibitory
self-interactions are not only responsible for the enhancement of
the fraction of flipping spins, a feature that is known from
Little's model, but even for the presence of fixed-point correlated
states as demonstrated by the lower part of the $D$ phase. The
presence of the various stationary states shown in this work depends
on the relationship between the self-interaction $J_0$, the initial
overlap $m_{\mu}^0$ and the value of $\nu$. These quantities shape
the basins of attraction of the simplest stationary states and other
states could be considered with alternative initial states if
necessary. An interesting extension of this work would be to
consider random self-interactions.

\ack

F. L. Metz thanks Prof. Desir\'e Boll\'e for the kind hospitality at the
Institute for Theoretical Physics of the Catholic University of
Leuven, where this work was completed, and acknowledges a fellowship
from CNPq (Conselho Nacional de Desenvolvimento Cient\'{\i}fico e
Tecnol\'ogico), Brazil. The work of one of the authors (WKT) was
financially supported, in part, by CNPq and a grant from FAPERGS
(Funda\c{c}\~ao de Amparo \`a Pesquisa do Estado de Rio Grande do
Sul), Brazil, to the same author is gratefully acknowledged.


\section*{References}

\begin{thebibliography}{24}

\bibitem{SK86} Sompolinsky H and Kanter I 1986 \PRL {\bf 57} 2861

\bibitem{CS92} Coolen A C C and Sherrington D 1992 \JPA {\bf 25}
5493

\bibitem{GTA93} Griniasty M, Tsodyks M V and Amit D J 1993 {\it Neural
Computation} {\bf 5} 1

\bibitem{CT94} Cugliandolo L F and Tsodyks M V 1994 \JPA {\bf 27} 741

\bibitem{WSC95} Whyte W, Sherrington D and Coolen A C C 1995 \JPA
{\bf 28} 3421

\bibitem{DCS98} D\"uring A, Coolen A C C and Sherrington D 1998 \JPA {\bf 31} 8607

\bibitem{KA98} Kitano K and Aoyagi T 1998 \JPA {\bf 31} L613

\bibitem{FKDO99} Fukai T, Kimoto T, Doi M and Okada M 1999 \JPA {\bf 32} 5551

\bibitem{LC94} Laughton S N and Coolen A C C 1994 \JPA {\bf 27} 8011

\bibitem{YYK01} Yong C, Yinghai W and Kongqing Y 2001 \PR E {\bf
63} 041901

\bibitem{AC01} Coolen A C C 2001 {\it Handbook of Biological
Physics IV: Neuro-Informatics and Neural Modeling} ed F Moss and S
Gielen (Amsterdam: Elsevier) p 619

\bibitem{LC95b} Laughton S N and Coolen A C C \PR E {\bf 51} 2581

\bibitem{KO02} Kawamura M and Okada M 2002 \JPA {\bf 35} 253

\bibitem{WT03} Theumann W K 2003 {\it Physica} A {\bf 328} 1

\bibitem{MKO03} Mimura K, Kawamura M and Okada M 2004 \JPA {\bf 37} 6437

\bibitem{UHO04} Uezu T, Hirano A and Okada M 2004 {\it J. Phys.
Soc. Japan} {\bf 73} 867

\bibitem{CZYZ08} Chen Y, Zhang P, Yu L, Zhang S 2008 \PR E {\bf 77}
016110

\bibitem{LC95} Laughton S N and Coolen A C C 1995 {\it J.
Stat. Phys.} {\bf 80} 375

\bibitem{Bu06} Buzsaki G {\it Rhythms of the Brain} (New York:
Oxford University Press)

\bibitem{Li74} Little W A 1974 {\it Math. Biosci.} {\bf 19} 101;
Little W A and Shaw G L 1978 {\it Math. Biosci.} {\bf 39} 281

\bibitem{FK87} Fontanari J F and K\"oberle R 1987 \PR A {\bf 36} 2475;
Fontanari J F and K\"oberle R 1988 {\it J. Phys. France} {\bf 49} 13;
Fontanari J F and K\"oberle R 1988 \JPA {\bf 21} L259

\bibitem{Fo88} Fontanari J F 1988 PhD Thesis University of S\~ao
Paulo (S\~ao Carlos, Brazil)

\bibitem{BB05} Boll\'e D and Busquets Blanco J 2005 {\it Eur. Phys. J.} B {\bf 47} 281

\bibitem{BEV06} Boll\'e D, Erichsen Jr R and Verbeiren T 2006
{\it Physica} A {\bf 368} 311

\bibitem{MT08} Metz F L and Theumann W K 2008 \JPA {\bf 41} 265001

\bibitem{MC88} Miyashita Y and Chang H S 1988 {\it Nature} {\bf 331} 68

\bibitem{Mi88} Miyashita Y 1988 {\it Nature} {\bf 335} 817

\bibitem{Br94} Brunel N 1994 {\it Network: Comput. in Neural Syst.} {\bf 5} 449

\bibitem{MAB03} Mongillo G, Amit D J and Brunel N 2003 {\it European
Journal of Neuroscience} {\bf 18} 2011

\bibitem {MT07} Metz F L and Theumann W K 2007 \PR E {\bf 75} 041907

\bibitem{DD78} De Dominicis C 1978 \PR B {\bf 18} 4913

\bibitem{EO92} Eissfeller H and Opper M 1992 \PRL {\bf 68} 2094;
Eissfeller H and Opper M 1994 \PR E {\bf 50} 709

\bibitem{Ve03} Verbeiren T 2003 PhD thesis K. U. Leuven (Leuven, Belgium)

\end{thebibliography}
\end{document}